\documentclass[12pt]{article}

\pdfoutput=1
\usepackage{slashed}
\usepackage{color, verbatim}
\usepackage{latexsym}
\usepackage{epsfig}
\usepackage{slashed}
\usepackage{amsmath,accents}

\usepackage{amssymb}
\usepackage{graphicx}
\usepackage{bm}

\usepackage[font={small}]{caption}

\usepackage{cite}
\usepackage{hyperref}

\definecolor{myred}{rgb}{0.7, 0, 0}
\definecolor{myblue}{rgb}{0, 0, 0.7}
\definecolor{mygreen}{rgb}{0.04, 0.7, 0.5}
\hypersetup{colorlinks,citecolor=myred,linkcolor=myblue,urlcolor=myblue,linktocpage=true}

\makeatletter

\@addtoreset{equation}{section}
\makeatother

\def\lsim{\mathrel{\raise.3ex\hbox{$<$\kern-.75em\lower1ex\hbox{$\sim$}}}}
\def\gsim{\mathrel{\raise.3ex\hbox{$>$\kern-.75em\lower1ex\hbox{$\sim$}}}}

\newcommand{\be}{\begin{equation}}
\newcommand{\ee}{\end{equation}}
\newcommand{\bea}{\begin{eqnarray}}
\newcommand{\eea}{\end{eqnarray}}

\newcommand{\gev}{\text{GeV}}

\begin{document}

\thispagestyle{empty}

\begin{center}

\begin{center}

{\LARGE\sf Higgsino Dark Matter in an economical Scherk-Schwarz setup}

\end{center}

\vspace{.5cm}

\textbf{
 Antonio Delgado$^{\,a}$, Adam Martin$^{\,a}$, Mariano Quir\'os$^{\,a,\,b}$
}\\

\vspace{1.cm}

${}^a\!\!$ {\em {Department of Physics, University of Notre Dame, 225 Nieuwland Hall \\ Notre Dame, IN 46556, USA
}}

${}^b\!\!$ {\em {Institut de F\'{\i}sica d'Altes Energies (IFAE) and BIST, Campus UAB\\ 08193, Bellaterra, Barcelona, Spain
}}

\end{center}

\begin{quote}\small
\textsc{Abstract:} 
We consider a minimal natural supersymmetric model based on an extra dimension with supersymmetry breaking provided by the Scherk-Schwarz mechanism. The lightest supersymmetric particle is a neutral, quasi-Dirac Higgsino and, unlike in previous studies, we assume that all Standard Model fields are propagating in the bulk.  The resulting setup is minimal, as neither extra matter, effective operators, nor extra $U(1)$ groups are needed in order to be viable. The model has three free parameters which are fixed by the Higgsino mass -- set to the range 1.1-1.2 TeV so it can play the role of Dark Matter, and by the requirements of correct electroweak breaking and the mass of the Higgs. After imposing the previous conditions we find a benchmark scenario that passes all experimental constrains with an allowed range for the supersymmetric parameters. In particular we have found gluinos in the range 2.0-2.1 TeV mass, electroweakinos and sleptons almost degenerate in the range 1.7-1.9 TeV and squarks degenerate in the range 1.9-2.0 TeV. The best discovery prospects are: i.) gluino detection at the high luminosity LHC ($\gtrsim 3\, \text{ab}^{-1}$),  and ii.) Higgsino detection at next-generation dark matter direct detection experiments. The model is natural, as the fine-tuning for the fixed values of the parameters is moderate mainly because supersymmetry breaking parameters contribute linearly to the Higgs mass parameter, rather than quadratically as in most models.

\end{quote}
\vfill
  
\newpage
\section{Introduction}
In spite of its experimental elusiveness, low-scale supersymmetry still (arguably) remains as the most complete and best motivated beyond the Standard Model (BSM) theory~\footnote{Although justice should be done to other BSM theories which also show themselves elusive with respect to experimental searches.}. On top of solving the naturalness problem, supersymmetric theories with $R$ parity conservation have naturally candidates for thermal Dark Matter (DM),
 the neutralinos. A number of recent works~\cite{Gelmini:2016emn,Arcadi:2017kky,Plehn:2017fdg,Krall:2017xij,Roszkowski:2017nbc,Kowalska:2018toh} have pointed out that,  out of the different neutralino spectra, a nearly pure Higgsino lightest supersymmetric particle (LSP) with a 1.1-1.2~TeV mass range remains as the most phenomenologically appealing candidate to DM~\footnote{We have encoded, in the given Higgsino mass interval, the theoretical uncertainty in the calculation of the thermal relic abundance $\Omega h^2$: i) There is a small mixing effect of the Higgsino and the wino which tends to increase the annihilation cross-section; ii) There is a small effect of the running of the Higgsino mass between the scale $1/R$, where the boundary conditions are set, and the tree level Higgsino mass scale $q_H/R$, by which the Higgsino mass tends to decrease as its beta function is positive.}
 The attractiveness of this scenario relies on the fact that its capability to reproduce the observed value of $\Omega h^2=0.1186\pm 0.0020$~\cite{Ade:2015xua} comes from the gauge interactions of the Higgsino multiplet alone and does not require a delicate mixture of different neutralino states (so-called `well-tempering')~\cite{ArkaniHamed:2006mb}. 

In this paper we will consider a very predictive low-scale supersymmetry breaking model with an LSP Higgsino. The paradigmatic mechanism of natural supersymmetry breaking is the Scherk-Schwarz (SS) twist of boundary conditions in a supersymmetric five-dimensional (5D) theory~\cite{Scherk:1978ta,Scherk:1979zr,Quiros:2003gg}. Due to the geometric nature of the SS mechanism, the supersymmetry breaking contributions to the Higgs mass are linear, instead of quadratic (as in gravity or gauge mediation). As a result, the fine tuning is proportional to a mass ratio~($\delta m/m$), instead of the square mass ratio~($\delta m^2/m^2$), and is thus significantly smaller. Moreover, because of the low-scale character of supersymmetry breaking, radiative corrections below the compactification scale are moderate, along with their corresponding contributions to the tuning. 

Unlike  previous studies~\cite{Pomarol:1998sd,Antoniadis:1998sd,Delgado:1998qr,Delgado:2016vib}, the considered model is a 5D supersymmetric theory with all matter and gauge fields in the bulk. It has three free parameters: the compactification scale $1/R$, the supersymmetry breaking masses proportional to a real parameter $q_R$ and a supersymmetric mass in the Higgs sector proportional to another real parameter $q_H$~\cite{Pomarol:1998sd,Antoniadis:1998sd,Delgado:1998qr}. The mass of the lightest KK states in the Higgsino sector is given by $q_H/R$, and the mass of the lightest KK sfermions propagating in the bulk is $q_R/R$. The mass of the Higgs sector depends on both parameters $q_R$ and $q_H$. 

We will fix $q_H$ so that the Higgsino is the DM particle, i.e.~the LSP with a mass of $1.1$ to $1.2$ TeV. The two other parameters, $q_R$ and $1/R$ can be set by demanding correct electroweak symmetry breaking and that the mass of the physical Higgs boson be $125\, \gev$. 
 We find that for certain values of the parameters in the $(q_R,1/R)$ plane, the (bulk propagating) stop sector is capable of radiatively triggering electroweak breaking -- much as it happens in the MSSM for high-scale supersymmetry breaking -- despite the fact our model has a low supersymmetry breaking scale. This is an advantage over other similar 5D SS constructions with stops localized at the $y=0$ brane, where higher dimensional operators localized at the $y=0$ brane~\cite{Dimopoulos:2014aua,Garcia:2015sfa} or extra triplets in the bulk~\cite{Delgado:2016vib}  are required in order to drive EWSB. Moreover, integrating out the top/stop sector, including all KK-modes, provides a threshold effect for the Higgs quartic coupling $\lambda$. After evolving $\lambda$ to the electroweak scale (by the SM radiative corrections) it is sufficiently large that the model accommodates a $125$ GeV mass for the HIggs. This is again an advantage over similar constructions with stops localized in the $y=0$ brane, where an extra $U(1)^\prime$~\cite{Dimopoulos:2014aua,Garcia:2015sfa} or singlets and/or triplets~\cite{Delgado:2016vib} had to be introduced to accommodate the physical value of the Higgs mass. The Higgs mass condition will also carve out contours in the $(q_R,1/R)$ plane, so that the phenomenologically interesting values of $q_R$ and $1/R$ are given by the intersection of the Higgs mass $(q_R,1/R)$ curves with the $(q_R,1/R)$ contours from correct electroweak breaking.

In short, the simultaneous conditions of a 1.1 to 1.2 TeV Higgsino, a 125 GeV Higgs, and correct electroweak breaking provide a discrete set of values for the three parameters $(q_R,q_H,1/R)$ -- this is a non-trivial statement since it was not guaranteed a priori that a viable solution would exist. Moreover, for the $q_H, q_R$, and $1/R$ values consistent with these conditions,  we find solutions that have spectra that are completely compatible with current LHC superpartner and direct dark matter searches~\cite{Aaboud:2018mna,Sirunyan:2017kqq}. The net result is a model with three free parameters $(q_R,q_H,1/R)$ that is able to reproduce the correct electroweak breaking, the correct Higgs mass, provide a viable DM candidate, and passes all experimental bounds.

The plan of this paper goes as follows. In Sec.~\ref{model} we will introduce in some detail the 5D model and its mass spectrum. In Sec.~\ref{ewsb} we will describe the conditions on electroweak symmetry breaking. In Sec.~\ref{Higgsmass} we will compute the threshold corrections to the Higgs quartic coupling and the physical value of the Higgs mass. In particular we will impose on the light Higgs a mass of 125 GeV,  according to experimental measurements. The spectrum, and some experimental prospects to detect it, is presented in Sec.~\ref{spectrum}. Finally some concluding remarks are postponed to Sec.~\ref{conclusion}.

\section{The model}
\label{model}
Our starting point is a flat, five-dimensional space where the fifth dimension $y$ is compactified on the orbifold $S^1/\mathbb{Z}_2$, with branes at the two fixed points $y=0,\pi R$. We are going to embed the SM into a supersymmetric model in 5D~\cite{Scherk:1978ta,Scherk:1979zr,Quiros:2003gg}.  Since the minimal ($N=1$) supersymmetry in 5D is the equivalent of $N=2$ in four-dimensions (4D), we have to incorporate new fields into every multiplet to satisfy this extended algebra. As a result of the orbifold compactification, one can decompose every $N=2$ multiplet into two $N=1$ 4D multiplets, each with a definite transformation with respect to the $\mathbb{Z}_2$ symmetry. In particular, (on-shell) vector multiplets in the bulk are $\mathbb V=(V_M,\Sigma,\lambda^i)\equiv(V_\mu,\lambda^1_L)^{+}\oplus(\Sigma+iV_5,\lambda^2_L)^{-}$ where $i=1,2$ transforms as a doublet of $SU(2)_R$ and the parities under $\mathbb Z_2$ for the two $N=1$ multiplets are specified by the $\pm$ superscripts.
Similarly there are two bulk Higgs hypermultiplets $\mathbb H^a=(H^a_i,\Psi^a)$, where the index $a=1,2$ transforms as a doublet of a global group $SU(2)_H$, and $\Psi^a$ are Dirac spinors. The parity, 
$\mathbb Z_2\equiv\left.\sigma_3\right|_{SU(2)_H}\otimes \gamma_5$, decomposition 
is $\mathbb H^2\equiv (H^2_2,\Psi^2_L)^+\oplus (H^2_1,\Psi^2_R)^-$ and 
$\mathbb H^1\equiv(H^1_1,\Psi^1_R)^+\oplus (H^1_2,\Psi^1_L)^-$. As such, the chiral multiplets $\mathcal H_2=(H^2_2,\Psi^2_L)$ and $\mathcal H_1=(H^{1\dagger}_1,\bar\Psi^1_R)$ have zero modes and play the role of the Higgs sector of the MSSM.

In Refs.~\cite{Pomarol:1998sd,Antoniadis:1998sd,Delgado:1998qr}, the Scherk-Schwarz mechanism~\cite{Scherk:1978ta,Scherk:1979zr,Quiros:2003gg} was used to break supersymmetry by means of a $U(1)_R\otimes U(1)_H$ symmetry. The mass spectrum one gets from this procedure depends on the charges $(q_R,q_H)$. In fact, only $q_R$ breaks supersymmetry; $q_H \ne 0$ generates a Higgsino mass $q_H/R$, thus providing a solution to the $\mu$ problem of the MSSM. More specifically, after SS supersymmetry breaking the mass eigenstates are:
\begin{itemize}
\item
Two  Majorana gauginos $\lambda^{(\pm n)}=(\lambda^{1(n)}_L\pm \lambda^{2(n)}_L)/\sqrt{2}$, with masses $|q_R\pm n|/R$. 
\item
Two Dirac Higgsinos $\tilde H^{(\pm n)}=(\Psi^{1(n)}\pm \Psi^{2(n)})/\sqrt{2}$, with masses $|q_H\pm n|/R$. 

\item
Two  Higgses $h^{(\pm n)}=\left[ H^{1(n)}_1+H^{2(n)}_2\mp (H^{1(n)}_2-H^{2(n)}_1\right]/2$, with masses \\ $|q_R-q_H\pm n|/R$. 
\item
Two  Higgses $H^{(\pm n)}=\left[ H^{1(n)}_1-H^{2(n)}_2\mp (H^{1(n)}_2+H^{2(n)}_1\right]/2$, with masses \\ $|q_R+q_H\pm n|/R$. 
 \end{itemize}
where positive ($+n$) and negative ($-n$) modes combine into whole towers with $n\in \mathbb Z$.

The main difference with respect to the scenario proposed in Refs.~\cite{Pomarol:1998sd,Antoniadis:1998sd,Delgado:1998qr}\footnote{In Ref.~\cite{Delgado:1998qr} the third generation of quarks and leptons was localized in the $y=0$ brane.} is that we will consider all matter fields propagating in the bulk. As such, matter fields must be represented by hypermultiplets, e.g. $\mathbb Q_L=(\widetilde Q,\widetilde Q^c,q)\equiv (\widetilde Q,q_L)^+\oplus(\widetilde Q^c,q_R)^-$ for the SM left-handed top quark, where only the even chiral multiplet $\mathcal Q_L=(\widetilde Q,q_L)$ admits a zero mode and $(\widetilde Q,\widetilde Q^c)^T$ transforms as a doublet of $SU(2)_R$. The SS supersymmetry breaking gives squared masses, equal to $(q_R\pm n)^2/R^2$, to the two complex scalars $Q^{(\pm n)}=(\widetilde Q^{(n)}\pm \widetilde Q^{c(n)})/\sqrt{2}$, which then become a whole tower of complex scalars with $n\in \mathbb Z$. Moreover, the SS breaking does not affect the tower $n\in \mathbb Z$ of ($SU(2)_R$ singlet) Dirac fermions $q^{(n)}$. Their KK modes instead have mass $|n|/R$, so that the zero mode is massless and can be identified with the left-handed SM top quark $q^{(0)}_L$ (without a Dirac partner). The same logic applies to every other SM fermion, e.g.~the SM right-handed quark, $\mathbb U_R=(\widetilde U,\widetilde U^c,u)\equiv (\widetilde U,u_R)^+\oplus(\widetilde U^c,u_L)^-$ with mass eigenstates $U^{(\pm n)}=(\widetilde U^{(n)}\pm \widetilde U^{c(n)})/\sqrt{2}$. Since we want to recover a chiral theory at the zero model level we are going to assume there are no masses in the bulk. 

Interactions among hypermultiplets are forbidden in the bulk, but are permitted on the branes where the symmetry of the theory is reduced to $N=1$ supersymmetry. We include the superpotential at the $y = 0$ brane:
\be
W=\left(\widehat h_t \,\mathcal Q_L \,\mathcal H_2 \,\mathcal U_R+\widehat h_b\,\mathcal Q_L \,\mathcal H_1 \,\mathcal D_R+\widehat h_\tau \mathcal L_L \mathcal H_ 1 \mathcal E_R\right)\ \delta(y),\ 
\label{eq:superpotential}
\ee
where $\widehat h_{b,\tau,t}$ are the 5D bottom, tau and and top Yukawa couplings. This superpotential will generate mass terms for the zero mode fermions once electroweak symmetry is broken. Supersymmetry demands that the Yukawa interactions present in Eq.~\eqref{eq:superpotential} are accompanied by several other interactions among the scalar superpartners. To see the full set of scalar interactions, we first integrate out the auxiliary fields. Neglecting the small effects of $\widehat h_{b,\tau}$, the quartic 4D potential can be written as~\cite{Mirabelli:1997aj}:
\begin{align}
V&=\widehat h_t^2\left(| \widetilde Q(0) H_2^2(0)|^2+|\widetilde Q(0)  \widetilde U(0)|^2+|\widetilde  U(0) H_2^2(0)|^2  \right)\delta(0)\nonumber\\
&+\widehat h_t\left(\widetilde  Q(0) H_2^2(0)\partial_y \widetilde  U^c(0)+
 \widetilde Q(0)  \widetilde U(0)\partial_y H_1^2(0)+
\widetilde  U(0) H_2^2(0)\partial_y \widetilde  Q^c(0)+h.c.    
\right),
\label{eq:potential}
\end{align}
where $\pi R\delta(0)\equiv \sum_n 1$ and $\widetilde Q(0), H^2_2(0)$, etc.~are the values of the wave functions for the entire KK tower of $Q, h-H$, etc.~on the brane. This potential depends on even fields and the derivative ($\partial_y$) of odd fields. The origin of these $\partial_y$ terms resides in the fact that  auxiliary components of off-shell 5D multiplets localized at the brane are given by the auxiliary field of the corresponding even component minus the derivative $\partial_y$ of the scalar field of the odd component~\cite{Mirabelli:1997aj}.

Working with mass eigenstates and pulling out normalization factors, the fields in Eq.~\eqref{eq:potential} become:
\begin{equation}
H_2^2(0)=\frac{1}{\sqrt{2\pi R}}(h-H),\quad  \partial_y  H^2_1(0)=\frac{1}{\sqrt{2\pi R}}(\widehat h-\widehat H)\end{equation}
\begin{equation}
H_1^1(0)=\frac{1}{\sqrt{2\pi R}}(h+H),\quad  \partial_y  H^1_2(0)=-\frac{1}{\sqrt{2\pi R}}(\widehat h+\widehat H)\end{equation}
\be
\widetilde Q(0)=\frac{1}{\sqrt{\pi R}}\,Q,\ \widetilde U(0)=\frac{1}{\sqrt{\pi R}}\,U,\quad
\partial_y \widetilde Q^c(0)=\frac{1}{\sqrt{\pi R}}\,\widehat Q,\ \partial_y \widetilde U^c(0)=\frac{1}{\sqrt{\pi R}\,}\widehat U 
\ee
where we have used $h, H, Q$, etc.~to stand for their corresponding tower of KK modes (or their derivatives) evaluated at $y = 0$. Explicitly,
\be
h\equiv\sum_n h^{(n)},\ H\equiv\sum_n H^{(n)},\quad \widehat h\equiv\sum_n\frac{(q_R-q_H+n)}{R}h^{(n)},\
\widehat H\equiv\sum_n \frac{(q_R+q_H+n)}{R}H^{(n)}
\label{eq:sumas}
\ee
and
\be
Q\equiv\sum_n Q^{(n)},\ U\equiv\sum_n U^{(n)},\quad \widehat Q\equiv\sum_n\frac{(q_R+n)}{R} Q^{(n)},\ \widehat U\equiv\sum_n\frac{(q_R+n)}{R} U^{(n)}.
\ee
With these definitions, the potential at $y=0$ becomes
\begin{align}
V=&h_t\left[ (h-H)Q\widehat U+(\widehat h-\widehat H)QU+(h-H)\widehat Q U +h.c.\right]\nonumber\\
+&h_t^2 \left[ |(h-H)Q|^2+|QU|^2+|(h-H)U|^2  \right]\pi R\, \delta(0),
\label{eq:potencialfinal}
\end{align}
where $h_t=\widehat h_t/\sqrt{2\pi^3R^3}$ is the (4D) SM top-Yukawa coupling. The other interaction from Eq.~\eqref{eq:superpotential} that we will need is the Yukawa coupling:
\be
\mathcal L_Y=\frac{\widehat h_t}{\sqrt{2\pi^3R^3}}(h-H)\bar q_L u_R+h.c.\equiv h_t\,(h-H)\bar q_Lu_R+h.c.
\ee
where $h - H$ represents the full tower of fields given by Eq.~(\ref{eq:sumas}), and, similarly, $q_{L}$ and $u_R$ stand for the coherent sums of fermionic states, e.g.~$u_{L,R}\equiv \sum_n u_{L,R}^{(n)}$.

With the superpotential set, the model is completely determined (apart from the SM couplings) by three different free parameters, $q_R$, $q_H$ and $1/R$. We will fix these parameters in the following way: first, we will require the lightest neutral Higgsino to be the dark matter. This not only sets the hierarchy $q_R>q_H$, as the lightest Higgsino needs to be the LSP, but requiring that the Higgsino achieves the correct relic abundance also fixes its mass, $q_H/R \simeq 1.1-1.2$ TeV~\cite{ArkaniHamed:2006mb,Kowalska:2018toh}. To set the other two parameters, we impose: i.) Correct electroweak symmetry breaking, and ii.) A physical Higgs boson mass of $125\, \gev$. Both of these conditions can be expressed as curves in the $(q_R, 1/R)$ plane, so our final parameter points will be given by the intersection of the $(q_R, 1/R)$  curves from the electroweak breaking requirement with those from the Higgs mass condition.

Before detailing the electroweak breaking and Higgs mass conditions, let us review the general spectrum given the hierarchy $q_R > q_H$: 
\begin{itemize}
\item
For scales below $q_H/R$ the theory is just the SM. For $q_R \le 1/2$, the SM Higgs
$\mathcal H\equiv h^{(0)}$. This mode has a (tree level) mass squared of $(q_R-q_H)^2/R^2$, so that tuning $q_R=q_H$, as we have done in previous work~\cite{Delgado:1998qr}, makes it massless. However, in this paper we will set the parameters by the condition of electroweak breaking and a 125 GeV Higgs. For parameter sets with $q_R \simeq q_H$ ($q_R < 1/2$), $h^{(0)}$ is light and identified with the SM Higgs. In this case, the mode $\mathcal H^\prime\equiv H^{(-1)}$ is identified as the second, `heavy' Higgs, with a mass squared equal to $(q_R+q_H-1)^2/R^2$. 

If we instead chose $q_R > 1/2$, the spectrum stays the same, but the identification of the lightest states shifts~\footnote{The $h^{(0)}$ and $H^{(-1)}$ masses are related by the symmetry $
q_R\to 1-q_R $.}. Specifically, $H^{(-1)}$ is lighter than $h^{(0)}$ and is identified with the SM Higgs, and the lightest sfermions and gauginos correspond to the $n=-1$ mode instead of the $n=0$ mode. For definiteness, we will assume $q_R < 1/2$ from now on.

Note that the particular case $q_R=q_H=1/2$ would make both $h^{(0)}$ \textit{and} $H^{(-1)}$ massless, in which case the Higgs sector would contain two light doublets, a configuration disfavored by present Higgs data except in the ``alignment limit''~\cite{Craig:2013hca,Carena:2013ooa,Haber:2017erd}. As we will see, we do not have to worry about the $q_R=q_H=1/2$ case as it 
does not fulfill the required conditions of electroweak breaking and correct value of the Higgs mass.

\item
For scales between $q_H/R$ and $q_R/R$ the only extra particles (on top of the heavy Higgses) are the Dirac Higgsinos with a mass $q_H/R$.
\item
Gauginos, sfermions  and the gravitino are degenerate at the mass $q_R/R$, although their masses show some splitting due to electroweak breaking contributions and radiative corrections in the 4D theory below the compactification scale. Therefore, between the scales $q_R/R$ and $1/R$ the theory resembles the MSSM.
\item
Finally for scales above $1/R$ the theory becomes 5D and all KK modes start to propagate. 
\end{itemize}

\section{Electroweak breaking}
\label{ewsb}

As explained earlier, in our 5D SS model the SM Higgs field $\mathcal H$ is identified with $ h^{(0)}$ (as we assume $q_R \le 1/2)$. While the spectrum contains a second ``heavy'' Higgs $\mathcal H^\prime$, identified with the mode $H^{(-1)}$,
 the large hierarchy between the SM Higgs and the heavy Higgs sector (including the KK modes) means we should immediately integrate out the heavy Higgses. The resulting low energy Higgs potential contains only one Higgs doublet $\mathcal H$, as in the SM:
\be
V=m^2|\mathcal H|^2+\lambda |\mathcal H|^4
\ee
This potential yields $m_{\mathcal H}^2=2\lambda v^2$, where $v=246$ GeV, $m_{\mathcal H}\simeq 125$ GeV. These inputs fix the numerical value of $m^2$ for correct EWSB to $m^2 =- (m_{\mathcal H}/\sqrt{2})^2\simeq -(88.4 \textrm{ GeV})^2$. Therefore, the condition of electroweak breaking boils down to imposing this number on the model and thereby selecting out viable values of the inputs $q_R$ and $1/R$.

As we have seen in the previous section, $m^2$ receives a tree level contribution $m_0^2$, as
\be
m^2_0=
(q_R-q_H)^2/R^2 \label{eq:m0}
\ee
This mass (squared) is positive definite so that a negative value of $m^2$ must be induced radiatively.
Radiative contributions coming from gauge interactions, computed in Ref.~\cite{Delgado:1998qr}, are positive and cannot trigger electroweak breaking. They are given by
 \be
 \Delta_g m^2=\frac{3g^2+g_Y^2}{192\pi^4}\left[9\Delta m^2(0)+3\Delta m^2(q_R\pm q_H)-6\Delta m^2(q_R)-6\Delta m^2(q_H)  \right],
 \ee
where the plus sign corresponds to the $h^{(0)}$ square mass, and the minus sign to that of $H^{(-1)}$, and
 \be
 \Delta m^2(q)=\frac{1}{2R^2}[Li_3(e^{2\pi i q})+h.c.],
 \ee
 where $Li_n(x)=\sum_{k=1}^\infty x^k/k^n$  are polylogarithm functions.
 
\begin{figure}[htb]
\centering
\includegraphics[width=15cm]{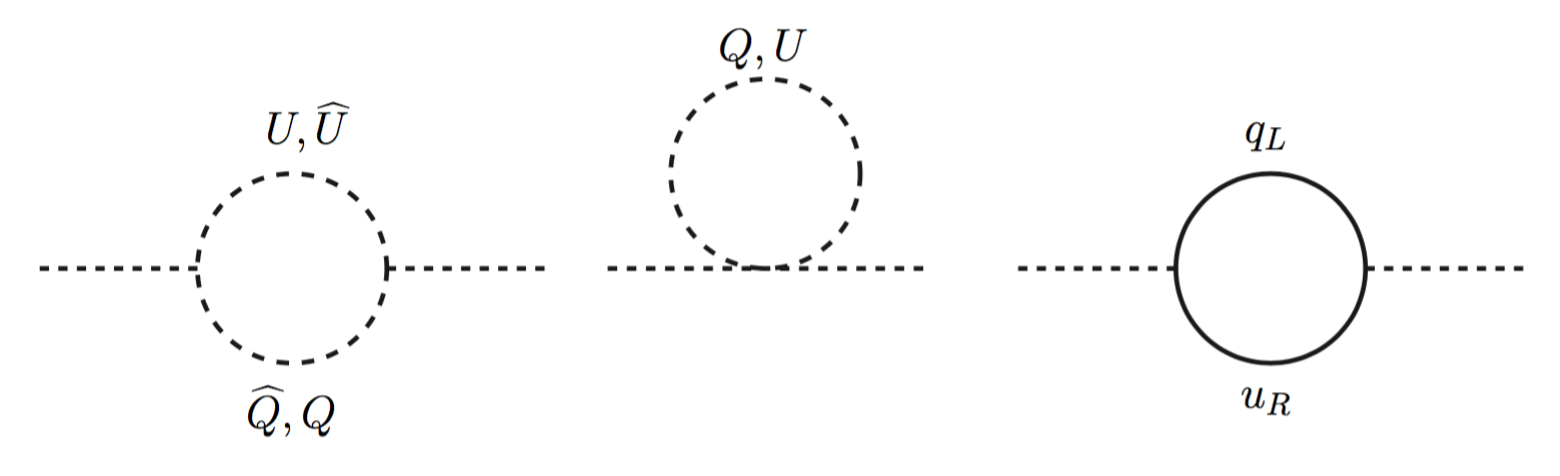} 
\caption{\it Diagrams proportional to $h_t^2$ contributing to $\Delta_t m^2_\mathcal H$. Note that in the loops full towers of bosons, $Q$, $U$, $\widehat Q$, $\widehat U$, and fermions $q_L$ and $u_R$, are exchanged.}
\label{fig:Deltam2}
\end{figure}

On the other hand, radiative corrections from the Yukawa coupling in Eq.~(\ref{eq:superpotential}), are negative and originate from the diagrams in Fig.~\ref{fig:Deltam2}.
The result is a finite, negative definite threshold correction that can be thought of as the result of integrating out the squark and quark  KK modes. The correction is  common for both $\mathcal H$ and $\mathcal H^\prime$, and is given by:
 \be
 \Delta_tm^2=\frac{3h_t^2(\mu)}{32\pi^4 R^2}\left[ 3 Li_3(e^{2\pi i q_R})-3i \cot(2\pi q_R) Li_4(e^{2\pi i q_R})-2\zeta(3)+h.c.
 \right].
 \label{eq:yukawaloop}
 \ee

To fix $q_R$ and $1/R$, we first match the high-energy and low-energy theories at the scale $\mu_0=q_R/R$, the scale where we integrate out the stop zero mode. In principle, the net effect of the matching is that the coupling in Eq.~\eqref{eq:yukawaloop} should be the top-Yukawa, evaluated with SM field content only, at $q_R/R$. In practice, the SM running of $m^2(\mu)$ between $m_t$ and $q_R/R$ has only a small effect on $m^2$ so we neglect it (this will not be the case when we examine the Higgs quartic $\lambda(\mu)$). With $\mu$ fixed to $q_R/R$ by the matching condition and $q_H/R$ fixed to the values between 1.1 and 1.2 TeV, the net tree-plus-loop Higgs mass is a function of $q_R$ and $1/R$ alone.

\begin{figure}[htb]
\centering
\includegraphics[width=10cm]{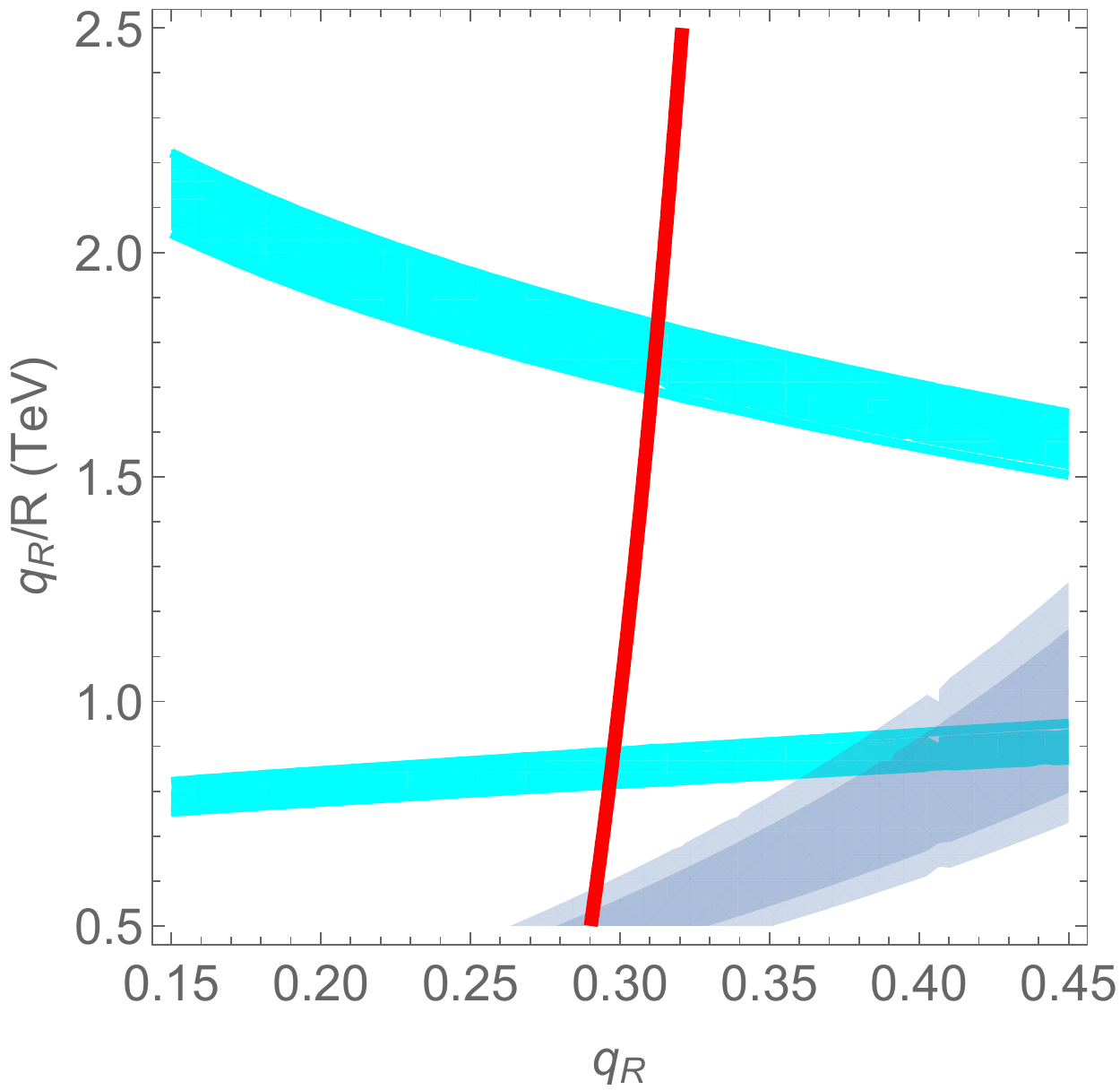} %
\caption{\it Cyan bands are the electroweak breaking conditions, for  $1.1 \textrm{ TeV}\leq q_H/R\leq 1.2 \textrm{ TeV}$. The shadowed gray areas are the corresponding (excluded) regions where $m^2_{\mathcal H'}<0$. The red thick solid line is the condition that $\lambda^{SM}=\Delta\lambda$ at the matching scale $\mu_0=q_R/R$.}
\label{fig:breaking}
\end{figure}

Setting now $m^2(q_R, 1/R)=-(88.4\textrm{ GeV})^2$, we can solve for the regions where EWSB is correctly achieved.  The result, plotted in the $(q_R,q_R/R)$ plane, is shown in Fig.~\ref{fig:breaking} below (red solid lines). As we vary the Higgsino mass $q_H/R$ between 1.1 - 1.2 TeV, the EWSB curves turn into a band ($q_H/R = 1/2$ TeV is the upper boundary, $1.1$ TeV is the lower one). For every case, we have to impose that the heavy Higgs $\mathcal H'$ is not tachyonic and heavy enough to justify our use of the decoupling limit~\footnote{The potential for the heavy Higgs is very steep. Therefore, we use the condition $m_\mathcal {H'}^2 > 0$ to approximate where the heavy Higgs is in the decoupling limit}. Regions where this condition, imposed on the full tree plus loop level $\mathcal H'$ mass, fail are shaded blue in Fig.~\ref{fig:breaking}.
 
Before moving on, there is one subtlety in the $\Delta m^2$ calculations that we would like to mention. The diagrams in Fig.~\ref{fig:Deltam2} exist between any two Higgs external states in the KK tower -- $n = 0$ in and $n = 0$ out, as well as for $n = 1$ in and $n= 0$ out, $n = 2$ in and $n= 0$ out, etc. As a result, the mass matrix for the KK Higgs states is not diagonal, with off diagonal entries all $\Delta_t m^2$ (in addition to $\Delta_t m^2$ contributions to the diagonal terms). Diagonalizing this mass matrix, the zero mode mass squared eigenvalue shifts by $\mathcal O[(\Delta_t m^2)^2R^2]$. However, we have checked that this effect is small and, as it is parametrically comparable to two-loop effects which we are not considering in this paper, therefore we will ignore it in our numerical calculations.

\section{The Higgs mass}
\label{Higgsmass}

\begin{figure}[htb]
\centering
\includegraphics[width=10cm]{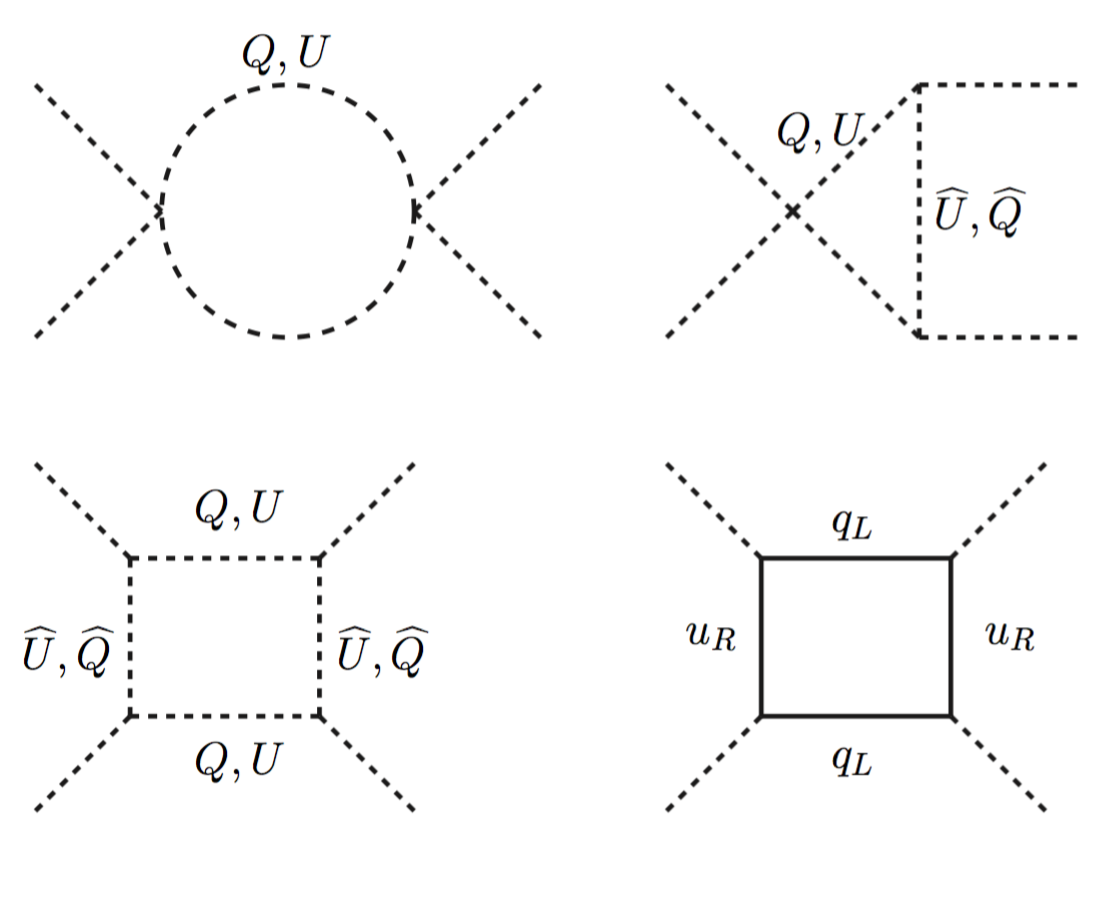} 
\caption{\it Diagrams proportional to $h_t^4$ contributing to $\Delta\lambda$.
}
\label{fig:Deltalambda}
\end{figure}

The second condition we impose to nail down $q_R$ and $1/R$ is a physical Higgs mass of 125 GeV. The Higgs mass in this theory is determined by the value of the quartic coupling $\lambda=\lambda_0+\Delta\lambda$, where $\lambda_0$ is the tree level value and $\Delta \lambda$ is the loop contribution. The quartic coupling at tree level vanishes, $\lambda_0=0$~\cite{Pomarol:1998sd}, so that the whole Higgs mass is controlled by the radiative piece. Here we will approximate $\Delta \lambda$ by the dominant contribution, coming from diagrams involving the top Yukawa. The relevant diagrams are shown in Fig.~\ref{fig:Deltalambda}, where as before $Q$, $U$, $\widehat Q$, $\widehat U$, $q_L$ and $u_R$ correspond to whole KK towers of states.
The final expression is given by the Euclidean momentum integral
\be
\Delta\lambda=\frac{3h_t^4(\mu)}{8\pi^2}\int_0^\infty p^7\left[s^4(p,0)-s^4(p,q_R)    \right]dp,
\label{eq:integral}
\ee
where we will fix $\mu$ to the matching scale and the function $s(p,q)$ is defined as
\be
s(p,q)=\frac{\pi R \sinh(2p\pi R)}{p[\cosh(2p\pi R)-\cos(2\pi q)]}.
\label{eq:tower}
\ee
As in the previous section, we will set the matching scale to $\mu_0 = q_R/R$ where the high energy and low energy theories coincide.
Notice that, for consistency, we have omitted any effect from the mixing among the heavy KK modes as this would correspond to a (small) two-loop effect. 

The integral (\ref{eq:integral}) is UV convergent. However, it has a logarithmic IR divergence originating from the contribution of the (massless) top quark zero mode in the loop. This divergence should be regularized by introducing an IR cutoff at the scale $m_t$.  A careful inspection reveals that
 $\Delta \lambda$, can be decomposed as 
\be
\Delta\lambda(\mu_0)=\Delta\lambda_{log}(\mu_0)+\Delta\lambda_{th}(\mu_0) 
\ee
where the IR cutoff and compactification scale are lumped together into one term:
\be
\Delta\lambda_{log}(\mu_0)=-\frac{3h_t^4(\mu_0)}{8\pi^2}\log(m_t R).
\label{eq:lambdarunning}
\ee
From the form of Eq.~\eqref{eq:lambdarunning}, we see that $\Delta\lambda_{log}$ is reminiscent of a similar contribution which appears in the MSSM, where the role of $\log (m_tR)$ is played by $\log(m_t/M_{SUSY})$~\cite{Haber:1996fp}. In our setup here, the scale of all contributions is fixed by the compactification scale $1/R$, which explains the origin of the argument of the logarithm in Eq.~(\ref{eq:lambdarunning}).

 The remaining term $\Delta\lambda_{th}$ has no explicit $1/R$ dependence and is therefore finite (as $R \to 0$). The only $1/R$, or $\mu$, dependence is implicit, in the scale where we evaluate the top Yukawa coupling $h_t(\mu)$ As such, $\Delta\lambda_{th}$ is a genuine threshold correction. The form of $\Delta \lambda_{th}$ is reminiscent of the stop threshold effect in the MSSM, with $q_R/R$ playing the role of the left-right mixing parameter $A_t$~\cite{Haber:1996fp,Carena:1995bx,Carena:1995wu}. This $q_R/R \to A_t$ identification can better understood by expanding out the first line of Eq.~(\ref{eq:potencialfinal}) and focusing on the zero modes. Among the terms, we find the trilinear $[3(q_R/R)-q_H/R]\mathcal H  Q_L^{(0)} (U_R^{(0)})^\ast+h.c.$.
 \begin{figure}[htb]
\centering
\includegraphics[width=7.9cm]{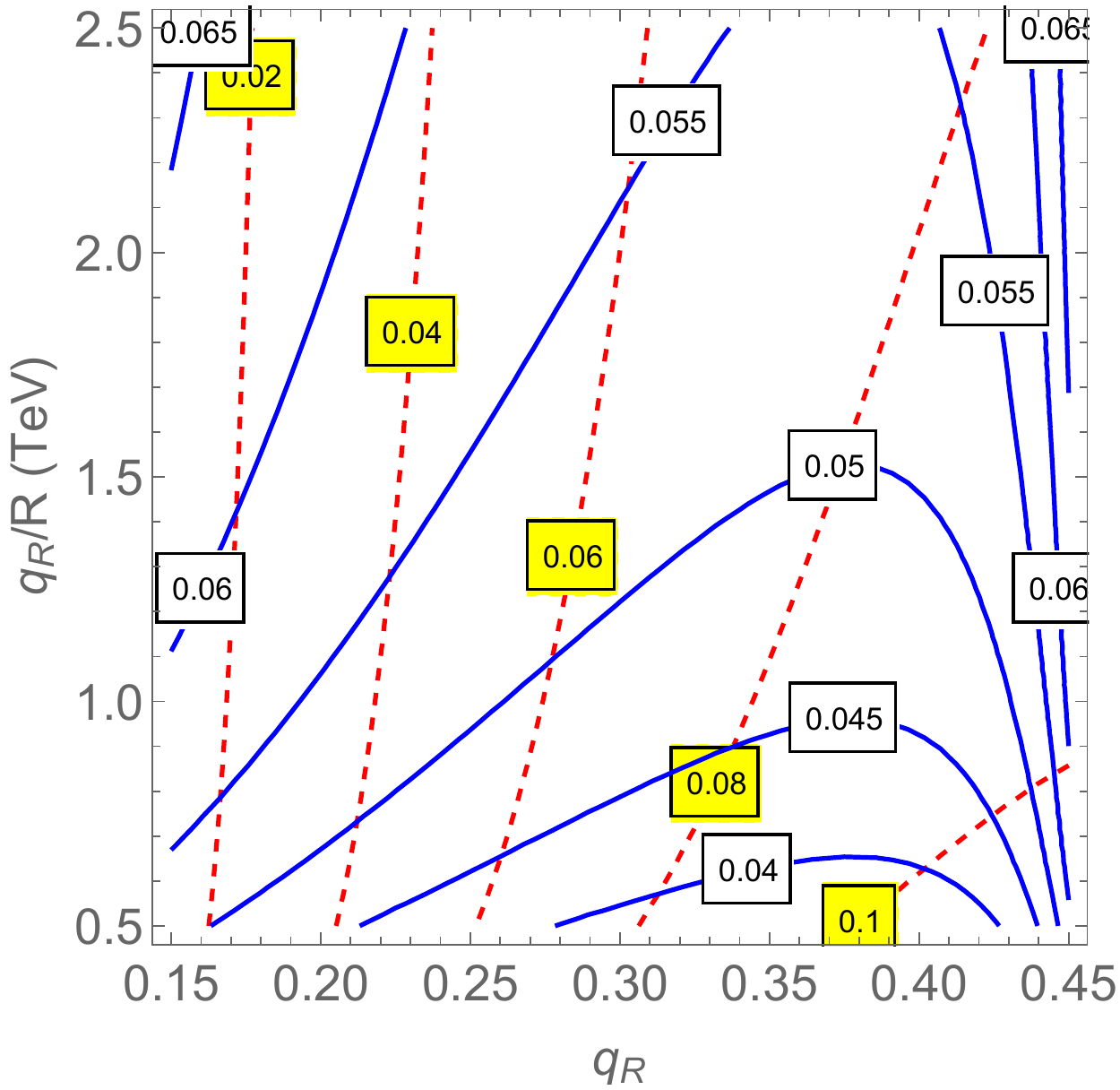}
\includegraphics[width=7.9cm]{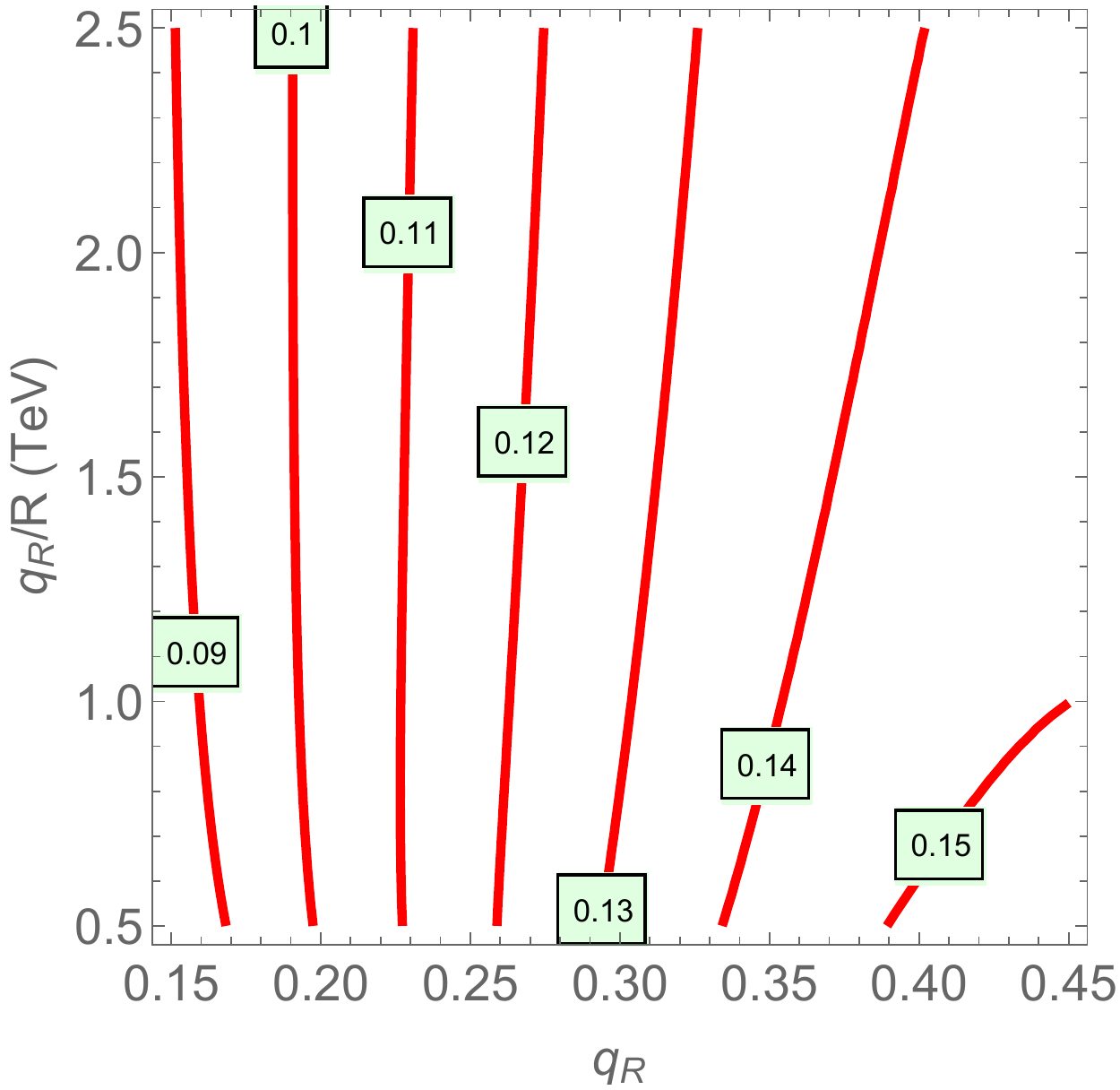} 
\caption{\it Left panel: Contour lines of $\Delta\lambda_{log}(\mu_0)$ (blue solid lines) and $\Delta\lambda_{th}(\mu_0)$ (red dashed lines) in the plane $(q_R,q_R/R)$ for $\mu_0=q_R/R$. Right panel: Contour lines of $\Delta\lambda(\mu_0)$ (red thick solid lines)  in the plane $(q_R,q_R/R)$.}
\label{fig:varios}
\end{figure} 

The relative weights of $\Delta\lambda_{log}(\mu_0)$ and $\Delta\lambda_{th}(\mu_0)$ are shown in the left panel or Fig.~\ref{fig:varios}. As expected, the value of $\Delta\lambda_{th}$ (red dashed lines) increases when $q_R$ increases (keeping fixed $q_R/R$), while $\Delta\lambda_{log}$ increases when $q_R$ decreases (for fixed $q_R/R$) as the value of $1/R$ increases. In this way, the effect is dominated by  $\Delta\lambda_{log}$ for low values of $q_R$, while for large values of $q_R$ it is dominated by $\Delta\lambda_{th}$. In the right panel of Fig.~\ref{fig:varios} we show contour lines of the total $\Delta\lambda$ (red solid lines). As we can see, the value of $\Delta\lambda$ increases with increasing $q_R$ as  the threshold corrections become more important in this region. 
 
 Note that our procedure is conservative. Had we fixed the matching scale at the top quark mass ($\mu = m_t$), we would have considered the renormalization group running of the quartic coupling between the scales $1/R$ and $m_t$ in the one-loop approximation. As shown in MSSM Higgs studies~\cite{Haber:1996fp,Carena:1995bx,Carena:1995wu}, one-loop running from the cut-off to $m_t$ overshoots the Higgs mass, yielding a larger value than the result if all large logarithms are resummed by renormalization group techniques.
 
 Finally, the matching condition is then given by~\footnote{As we have already integrated out the top quark in the contribution of the corresponding tower to Eq.~(\ref{eq:tower}), and in the approximation we are considering where we are neglecting the contribution from the gauge $g$ and $g_Y$, and quartic $\lambda$, couplings, $\lambda^{SM}$ is given by its value at the scale $m_t$.}:
\be
\Delta\lambda(\mu_0)=\lambda^{SM}=\frac{m_{\mathcal H}^2}{2v^2}
\label{eq:finalHH}
\ee
Using this condition (\ref{eq:finalHH}) with $m_{\mathcal H}$ and $v$ fixed to their experimental values sets $q_R$ and $1/R$. The allowed values in the $(q_R, q_R/R)$ plane are shown in Fig.~\ref{fig:breaking}. The near verticality of solid lines reflects the $1/R$ independence of  $\Delta\lambda/h_t^4$. 
\section{The spectrum and phenomenology}
\label{spectrum}
As we can see from Fig.~\ref{fig:breaking} the $(q_R,q_R/R)$ points that satisfy both the EWSB and $m_\mathcal H = 125\, \gev$ conditions correspond to the intersection between the cyan bands and the red solid line. There are two intersecting regions, but only only one of them remains if we impose that the Higgsino be the LSP, or in other words if we restrict ourselves to the region for which $q_R/R>q_H/R$. The bounds on the parameter values for the remaining region, as well as some details of the spectra are listed in Tab.~\ref{table}. Parameters in the first (second) row correspond to the lower (upper) endpoints which correspond to the value $q_H/R=1.1$ TeV ($q_H/R=1.2$ TeV). Notice that, as observed earlier, to avoid multiply repeated solutions we have restricted ourselves to $q_R<1/2$.
\begin{table}[h!]
\begin{center}
\begin{tabular}{||c|c|c|c|c|c|c||}
\hline\hline
 $q_R$ & $q_H$ & $1/R$ (TeV) & $q_R/R$ (TeV) &$q_H/R$ (TeV) & $M_{\tilde g}$ (TeV)&$m_{\mathcal  H'}$ (TeV) \\
\hline
0.31 & 0.2 & 5.5 & 1.7 &1.1&2.0& 2.7\\
 0.31&0.2&5.9&1.9&1.2&2.1&2.9\\

\hline\hline
\end{tabular}
\caption{\it Range from Fig.~\ref{fig:breaking} that satisfies the conditions of correct electroweak breaking for a single Higgs field, the correct value of the Higgs mass at 125 GeV, and the Higgsino with a mass in the range between 1.1 and 1.2 TeV being the LSP. The supersymmetric parameters for 
the first (second) row is the lower (upper) limit of the range.}
\label{table}
\end{center}
\end{table}

Some comments about the spectrum: 
\begin{itemize}
\item
The mass of the neutral ($\chi^0_1, \chi^0_2$) and charged ($\chi^{\pm}$) Higgsinos is given by $q_H/R$ and has been fixed to be in the range $1.1 \textrm{ TeV}<q_H/R<1.2$ TeV. The lightest neutral Higgsino is the LSP for all the range, and is the dark matter. In the unbroken phase, both the charged and neutral components of the Higgsino are Dirac fermions. After electroweak symmetry breaking,  the two neutral Higgsinos become Majorana fermions split by 
$\delta\simeq 13$ GeV. Similarly, mixing with the wino shifts the charged Higgsino mass such that 
$m_{\chi^\pm}-m_{\chi_1^0}\simeq 3$ GeV. Furthermore, there is a radiative correction to the mass of the charged Higgsino as $\Delta \sim 340$ MeV~\cite{Cirelli:2005uq} such that the mass difference between the lightest chargino and the neutral LSP is $\sim 3.5$ GeV.
\item
In the scalar Higgs sector we have the SM Higgs $\mathcal H$, whose mass has been fixed to the experimental value 125 GeV, and a heavy inert Higgs doublet $\mathcal H^{\prime}$. The four scalars in the inert doublet -- one CP-even,  one CP-odd, two charged -- are degenerate and with a mass in the range 2.7 TeV$<m_{\mathcal H^\prime}<$ 2.9 TeV. 
Of course, there is a plethora of other inert scalar doublets corresponding to the rest of the tower of KK modes; however they are all heavier than (or of order of) $1/R$ and thus they cannot be consistently included in the 4D effective theory. In short, all heavy Higgs doublets are heavier than the Higgsino, thus the spectrum below $q_H/R$ is the \textit{pure} Standard Model one. 
\item
All sfermion $m_{\tilde f}$  and gaugino $M_a$ $ (a=1,2,3)$ tree-level masses are degenerate to the value of $q_R/R$ in the range 1.7 TeV$< q_R/R<$ 1.85 TeV at the supersymmetry breaking scale $1/R$.  The degeneracy will be lifted by the 4D renormalization group running between the scale $1/R$ and $q_R/R$.  
In particular, the running gluino mass will be increased between $1/R$ and $q_R/R$ by $\sim 5 \%$ so that the value of the running gluino mass $M_3(q_R/R)$ will be in the range between 1.8 TeV and 1.9 TeV. Moreover radiative corrections relating the running gluino mass ($M_3$) with the pole gluino mass ($M_{\tilde g}$) amount, for the range of approximately equal squark and gluino masses, to an increase of the gluino mass by $\sim 10\%$~\cite{Martin:2006ub}, leading to the range 2.0 TeV$\lesssim M_{\tilde g}\lesssim 2.1$ TeV, as shown in Tab.~\ref{table}.
Similarly, squarks obtain one-loop QCD radiative corrections between $1/R$ and $q_R/R$, which increase their squared mass by $\sim 5\%$. The final value of the running squark mass will be in the range between $\sim$ 1.8 TeV and $\sim$ 1.9 TeV, while the corrections leading to squark pole masses ($M_{\tilde q}$) amount to $\sim 5\%$~\cite{Martin:2006ub} leading to the range between 1.9 TeV and 2.0 TeV. The slepton masses also receive radiative corrections, though they are much smaller so the one-loop results are essentially $\sim q_R/R$.
\item
In all cases, and typical of the Scherk-Schwarz mechanism, the gravitino mass is
\be
m_{3/2}=q_R/R
\ee

\end{itemize}

Current searches of gluinos put a bound on their mass around $2$ TeV for the case of a $1.1$ TeV neutralino~\cite{Aaboud:2017hrg,Sirunyan:2017cwe}. However, that analysis assumes a 100\% branching rate~\footnote{$\tilde g \to \bar b b + \slashed E_T$ is  the most sensitive gluino decay channel. All other channels were considered in Ref.~\cite{Aaboud:2017hrg,Sirunyan:2017cwe}, each with 100\% branching fraction.} $\tilde g \to b\bar b + \chi^0_1$, while the degeneracy of the squark spectrum in our model means $BR(\tilde g \to b\bar b + \chi^0_1) \sim 1/6$. Naively recasting the excluded cross section from~\cite{Aaboud:2017hrg,Sirunyan:2017cwe} into a `democratically decaying gluino' scenario, we find the gluino mass bound relaxes to $1.7$ TeV for a LSP of $1.1$ TeV and essentially no bound for a $1.2$ TeV LSP. Taking into account that the pole mass of the gluino in our model is 
in the rage between 2.0 TeV and 2.1 TeV we find that our model is safe from current LHC searches.

The model may be probed at the HL-LHC and in future colliders. The best place to look for a signal at the LHC will be in classic gluino pair production channels: $pp \to \tilde g\tilde g \to 2q\,2\bar q + 2\chi $. Of these, gluinos that decay into third generation quarks plus missing energy, $\tilde g \tilde g \to 2b 2\bar b+ \slashed E_T, 2t 2\bar t + \slashed E_T, b \bar b t\bar t + \slashed E_T$, etc. provide the most experimental handles (third generation tags, leptons from $t$ decay, etc) and should be the most effective. Ref.~\cite{Shchutska:2016giq} explored a spectrum similar to ours and showed that the reach of the HL-LHC would be around $2.5$ TeV (after $3$ ab$^{-1}$) for our neutralino mass range in the top plus missing energy channel, though as in Ref.~\cite{Aaboud:2017hrg} this limit came from assuming gluinos decay $100\%$ of the time to only one quark-squark flavor.

Of special interest is the LSP Higgsino, in the range between $1.1$ and $1.2$ TeV. Higgsino LSPs are best probed at colliders through the production of their chargino and heavier neutralino cousins $pp \to \chi^{\pm}\chi^0_2$, which subsequently decay back to the LSP $\chi^0_1$. However, in our setup the entire Higgsino multiplet is highly degenerate, $m_{\chi^{\pm}} - m_{\chi_1^0} \lesssim 4\, \gev$. A $\sim 4\, \gev$ splitting is sufficiently large that the decays will be prompt and therefore  techniques based on displaced vertices~\cite{Feng:1999fu,Gunion:1999jr,CMS:2014wda,Mahbubani:2017gjh,Liu:2018wte} do not apply. At the same time, a $4\, \gev$ splitting is small enough that the particles emitted $\chi^{\pm}, \chi^0_2$ decays to the LSP are too soft to pass triggering and identification requirements. If the entire chargino/neutralino system is boosted, the decay products inherit this boost and can be pushed above trigger/identification thresholds, though a large boost requires a hard object for the chargino/neutralino system to recoil against and significantly decreases the production rate~\cite{Giudice:2010wb,Schwaller:2013baa, Han:2014kaa, Han:2014xoa}. The net result is that 1.1 to 1.2 TeV Higgsino discovery at the LHC is essentially impossible due to the small cross-section of sufficiently boosted Higgsino pairs. However, at a collider with more energy, the signal cross section is higher and it may be possible to create sufficient amounts of highly boosted charginos/neutralinos for discovery~\cite{Bramante:2014tba,Bramante:2014tba2, Low:2014cba}.

A better option for discovering the Higgsino LSP is via dark matter direct detection experiments~\cite{Bramante:2014tba,Bramante:2014tba2}. The detection prospects depend strongly on the bino/wino admixture in the LSP, as that admixture controls the strength of the LSP-LSP-Higgs vertex that drives spin-independent scattering rate off nuclei~\footnote{Higgsino LSPs can be detected via their spin-dependent scattering off nuclei, although the prospects there are not as good~\cite{Kowalska:2018toh}.}. The LSP for our benchmark points is around $99\%$ pure Higgsino  --  a result of the large wino/bino mass --  so the spin-independent nuclear cross section for the benchmark range is $\sim 10^{-10}$ pb~\cite{Kowalska:2018toh}; thus, the whole range escape the current limit from XENON-1T~\cite{Aprile:2017iyp}. However, as shown in Ref.~\cite{Kowalska:2018toh}, Higgsino LSPs of this purity will be accessed in the next generation experiments like XENON-nT or LZ.

\section{Conclusion}
\label{conclusion}

In this paper, we have presented an economical, predictive supersymmetric model where (5D) supersymmetry is broken by the Scherk-Schwarz mechanism. The model has three free parameters: $q_R$, $q_H$ and $1/R$; or, equivalently the mass of gauginos and sfermions, $q_R/R$, the Higgssino mass $q_H/R$ and the KK mass $1/R$. The conditions for electroweak breaking, and a physical Higgs boson mass of 125 GeV fix $q_R/R$ and $1/R$, while the Higgsino mass is set in the range between
$1.1$ TeV and $1.2$ TeV to reproduce the observed DM relic abundance (thus fixing $q_H/R$). We find a range in the parameters space that can reproduce the aforementioned conditions, which correspond to sparticle masses in the range between $\sim 1.7$ TeV (for $q_H/R=1.1$ TeV) and  $\sim 1.9$ TeV (for $q_H/R=1.2$ TeV). By considering that the corresponding range of pole gluino masses is between 2.0 TeV and 2.1 TeV, the whole range of points seem to pass all experimental bounds. 

Moreover there is no chance of detecting the Higgsino LSP at the LHC as the neutralino/chargino components are highly degenerate. The best chance for discovery is instead at next-generation direct detection experiments for dark matter like XENON-nT or LZ. The LHC prospects for the gluino are better, as studies of similar spectra project sensitivity to $\sim 2.5\, \text{TeV}$ after an integrated luminosity around 3 ab$^{-1}$ at the HL-LHC.

One of the main features of the SS supersymmetry breaking mechanism is that masses $m_i=\{q_R/R,q_H/R\}$ contributing to the Higgs mass term $m_0^2|\mathcal H|^2$, are added linearly as in Eq.~(\ref{eq:m0}),
and not quadratically as in other mechanisms of supersymmetry breaking such as gravity or gauge mediation. As a consequence of this linear behavior, the fine tuning according to  the sensitivity definition~\cite{Barbieri:1987fn}
\be
\Delta_i=\left|\frac{\partial \log m_0^2}{\partial \log m_i^2 }\right|=\left|\frac{\partial \log m_0}{\partial \log m_i}\right|=\frac{m_i}{m_0}
\ee
 is milder. In particular for the benchmark points of Tab.~\ref{table} the fine-tuning among scales $q_R/R$ and $q_H/R$ is around $(q_R-q_H)/q_R\sim 0.3$, while for conventional mechanisms where contributions are added quadratically, the tuning would be $\lesssim 1\%$. 

Finally, while we have focused on the scenario where the Higgsino mass is set from the outset, it is worth considering what happens if we drop this requirement. It is surprising that the present bounds on the gluino mass are set by Higgsino masses $\gtrsim \mathcal O(1)$ TeV, in the ballpark where they reproduce the required thermal relic density. For smaller Higgsino masses a second dark matter component would be needed. By the same logic, increasing the Higgsino mass while maintaining its role as the LSP, the annihilation cross section is too weak and the thermal Higgsino relic density would overclose the universe. One way to make this heavier scenario viable is to introduce a small source of R-parity breaking --small enough to be consistent with collider bounds but enough to make the Higgsino decay in the early universe. Of course, in that case, an alternative candidate to dark matter should be provided by the theory.

\section*{Acknowledgments}
The work of AD and AM is partly supported by the National Science Foundation under grant PHY-1820860. The work of MQ is partly supported by Spanish MINEICO (grants CICYT-FEDER-FPA2014-55613-P and FPA2017-88915-P), by the Catalan Government under grant 2017SGR1069, and Severo Ochoa Excellence Program of MINEICO (grant SEV-2016-0588).

\end{document}